\documentclass[preprint,12pt,a4paper]{elsarticle}

\usepackage{amssymb}
\usepackage{hyperref}
\usepackage{subcaption}
\usepackage{xcolor}
\usepackage{listings}
\usepackage{soul}

\usepackage{todonotes}
\usepackage{comment}
\usepackage{xspace}

\colorlet{punct}{red!60!black}
\definecolor{background}{HTML}{EEEEEE}
\definecolor{delim}{RGB}{20,105,176}
\colorlet{numb}{magenta!60!black}

\lstdefinelanguage{json}{
    basicstyle=\normalfont\ttfamily,
    numbers=left,
    numberstyle=\scriptsize,
    stepnumber=1,
    numbersep=8pt,
    showstringspaces=false,
    breaklines=true,
    frame=lines,
    backgroundcolor=\color{background},
    literate=
     *{0}{{{\color{numb}0}}}{1}
      {1}{{{\color{numb}1}}}{1}
      {2}{{{\color{numb}2}}}{1}
      {3}{{{\color{numb}3}}}{1}
      {4}{{{\color{numb}4}}}{1}
      {5}{{{\color{numb}5}}}{1}
      {6}{{{\color{numb}6}}}{1}
      {7}{{{\color{numb}7}}}{1}
      {8}{{{\color{numb}8}}}{1}
      {9}{{{\color{numb}9}}}{1}
      {:}{{{\color{punct}{:}}}}{1}
      {,}{{{\color{punct}{,}}}}{1}
      {\{}{{{\color{delim}{\{}}}}{1}
      {\}}{{{\color{delim}{\}}}}}{1}
      {[}{{{\color{delim}{[}}}}{1}
      {]}{{{\color{delim}{]}}}}{1},
}

\newtoggle{finalPaper}

\togglefalse{finalPaper} % version with annotations
\iftoggle{finalPaper} {

	\newcommand{\rmvtxt}[1]{}}
{

	\newcommand{\rmvtxt}[1]{\st{#1}}}

\journal{SoftwareX}

\newcommand{\DISINFOX}{{\sffamily DISINFOX}\xspace}

\begin{document}
\renewcommand{\labelenumii}{\arabic{enumi}.\arabic{enumii}}

\begin{frontmatter}

%% Title, authors and addresses

%% use the tnoteref command within \title for footnotes;
%% use the tnotetext command for theassociated footnote;
%% use the fnref command within \author or \address for footnotes;
%% use the fntext command for theassociated footnote;
%% use the corref command within \author for corresponding author footnotes;
%% use the cortext command for theassociated footnote;
%% use the ead command for the email address,
%% and the form \ead[url] for the home page:
%% \title{Title\tnoteref{label1}}
%% \tnotetext[label1]{}
%% \author{Name\corref{cor1}\fnref{label2}}
%% \ead{email address}
%% \ead[url]{home page}
%% \fntext[label2]{}
%% \cortext[cor1]{}
%% \address{Address\fnref{label3}}
%% \fntext[label3]{}

\title{DISINFOX: an open-source threat exchange platform serving intelligence on disinformation and influence operations}

%% use optional labels to link authors explicitly to addresses:
%% \author[label1,label2]{}
%% \address[label1]{}
%% \address[label2]{}

\author[label1]{Felipe Sánchez González}
\author[label2]{Javier Pastor Galindo\corref{cor1}}
\author[label1]{José A. Ruipérez-Valiente}
\address[label1]{Department of Information and Communications Engineering, University of Murcia, Spain}
\address[label2]{Department of Computer Systems Engineering, Universidad Politecnica de Madrid, Spain}

\cortext[cor1]{
    Corresponding author. \\ Email address: javier.pastor.galindo@upm.es (Javier Pastor-Galindo).}

\begin{abstract}
%% Text of abstract 
This paper introduces DISINFOX, an open-source threat intelligence exchange platform for the structured collection, management, and dissemination of disinformation incidents and influence operations. Analysts can upload and correlate information manipulation and interference incidents, while clients can access and analyze the data through an interactive web interface or programmatically via a public API. This facilitates integration with other vendors, providing a unified view of cybersecurity and disinformation events. 

The solution is fully containerized using Docker, comprising a web-based frontend for user interaction, a backend REST API for managing core functionalities, and a public API for structured data retrieval, enabling seamless integration with existing Cyber Threat Intelligence (CTI) workflows. In particular, DISINFOX models the incidents through DISARM Tactics, Techniques, and Procedures (TTPs), a MITRE ATT\&CK-like framework for disinformation, with a custom data model based on the Structured Threat Information eXpression (STIX2) standard.  

As an open-source solution, DISINFOX provides a reproducible and extensible hub for researchers, analysts, and policymakers seeking to enhance the detection, investigation, and mitigation of disinformation threats. The intelligence generated from a custom dataset has been tested and utilized by a local instance of OpenCTI, a mature CTI platform, via a custom-built connector, validating the platform with the exchange of more than 100 disinformation incidents.
\end{abstract}

\begin{keyword}
%% keywords here, in the form: keyword \sep keyword
Disinformation \sep Influence Operations \sep Foreign Information Manipulation and Interference (FIMI) \sep Cybersecurity \sep Cyber Threat Intelligence (CTI) \sep DISARM framework
%% PACS codes here, in the form: \PACS code \sep code

%% MSC codes here, in the form: \MSC code \sep code
%% or \MSC[2008] code \sep code (2000 is the default)

\end{keyword}

\end{frontmatter}

%\linenumbers

\section*{Metadata}
\label{metadata}

\begin{table}[!h]
\begin{tabular}{|l|p{6.5cm}|p{6.5cm}|}
\hline
\textbf{Nr.} & \textbf{Code metadata description} & \textbf{Please fill in this column} \\
\hline
C1 & Current code version & v1.0 \\
\hline
C2 & Permanent link to code/repository used for this code version & \url{https://github.com/CyberDataLab/disinfox} \\
\hline
C3  & Permanent link to Reproducible Capsule & - \\
\hline
C4 & Legal Code License   & \href{https://github.com/CyberDataLab/disinfox/blob/main/LICENSE}{MITLicense} \\
\hline
C5 & Code versioning system used & git \\
\hline
C6 & Software code languages, tools, and services used & Docker, Python, Flask, jinja2, HTML, Bootstrap, JavaScript \\
\hline
C7 & Compilation requirements, operating environments \& dependencies & Windows/Ubuntu/MacOS, Docker\\
\hline
C8 & If available Link to developer documentation/manual & \url{https://github.com/CyberDataLab/disinfox/blob/main/README.md} \\
\hline
C9 & Support email for questions & felipe.sanchezg@um.es \\
\hline
\end{tabular}
\caption{Code metadata}
\label{codeMetadata} 
\end{table}

\section{Motivation and significance}

Disinformation has shaped public opinion throughout history  \cite{posetti2018short}.  
However, its impact has dramatically increased with the rise of digital communication and the unprecedented speed at which false narratives can spread \cite{spreadfalseinformation}.  
For example, the Ukrainian war has demonstrated that disinformation attacks are not merely side effects of conflicts but key strategic components in modern geopolitical warfare \cite{ukrdisinfoevolution}.  
As a result, detecting, analyzing, and sharing disinformation intelligence has become a priority among policymakers and security organizations \cite{eeas1, eeas2}.  

Disinformation incidents are often launched alongside traditional cyberattacks to amplify their impact, destabilize societies, and manipulate public perception~\cite{10492674}.  
Both cyberattacks and disinformation campaigns frequently exploit the same digital channels, making them deeply interconnected threats.  
Thus, disinformation can be considered a cybersecurity concern, as it complements cyber operations by influencing public opinion, disrupting critical systems, and obscuring the truth behind digital intrusions~\cite{disinfoiscybersecurity, baraniuk2024potential}.  

To combat the rise of cybersecurity threats, Cyber Threat Intelligence (CTI) was established to understand the capabilities, intent, motivations, and tactics, techniques, and procedures (TTPs) of adversaries \cite{10117505}.  
CTI feed platforms such as \textit{AlienVault OTX} and \textit{ThreatFox} aggregate reports and Indicators of Compromise (IoCs) from cybersecurity incidents worldwide using standardized formats like STIX (Structured Threat Information eXpression) and TAXII (Trusted Automated Exchange of Intelligence Information) to enable programmatic data extraction to endpoint CTI platforms such as \textit{OpenCTI} and \textit{MISP}, which integrate these feeds to enhance organizational awareness, improving both proactive and retrospective cybersecurity defenses.  

While CTI exchange platforms are highly mature in handling conventional cybersecurity incidents, they do not natively support disinformation incidents.  
Current efforts to monitor and analyze disinformation rely primarily on fact-checking repositories such as \textit{EUvsDisinfo}, which document disinformation campaigns in unstructured, natural language reports.  
This approach limits automated processing, making large-scale analysis and correlation across incidents nearly infeasible.  

Recently, frameworks and ontologies for structuring disinformation incidents have emerged \cite{disinfomodel, disinfomodel2, tudela2025influenceoperationontologyioo}.  
Specifically, the \textit{DISARM framework} \cite{disarm} provides a structured, MITRE ATT\&CK-like matrix for disinformation TTPs, offering a direct mapping to STIX objects.  
This enables a systematic approach to categorizing disinformation incidents and identifying the strategies used by malicious actors to manipulate public perception in digital spaces \cite{pastorgalindo2025influenceoperationssocialnetworks}.  

To address the absence of a dedicated threat exchange for disinformation, this work introduces \DISINFOX (DISINFOrmation threat eXchange), an open-source threat exchange platform for managing and sharing disinformation incidents in an interoperable format.  
\DISINFOX provides a dedicated repository with a custom data model based on DISARM TTPs where disinformation incidents can be structured, stored, and seamlessly integrated with traditional CTI solutions.  

This structured approach to disinformation intelligence offers several advantages.  
First, it enables automated processing, facilitating faster detection, preservation, and analysis of disinformation incidents.  
Second, by placing disinformation within the same analytical landscape as conventional cyber threats, \DISINFOX enhances the correlation of influence operations with traditional cyberattacks, revealing deeper threat patterns.  
To achieve this, \DISINFOX implements a modular, containerized system consisting of:  
(i) a web-based frontend for submitting incidents with automatic TTPs extraction and conducting preliminary analysis with interactive knowledge graphs and statistics,  
(ii) a RESTful backend API managing the platform’s core functionalities independently from the user interface,  
and (iii) a public API for programmatic access to newly uploaded disinformation incidents, enabling seamless integration with other CTI platforms.  

\section{Software description}

\subsection{Software architecture}

DISINFOX has been designed through a service-oriented architecture to maximize interoperability while maintaining scalability and modularity. The publicly available implementation\footnote{\url{https://github.com/CyberDataLab/disinfox}} relies on Docker containers for each service.
Docker containers are a light-weight virtualization technology that provides isolation for processes while reducing the resource usage of traditional virtual machines \cite{PASTORGALINDO202467}.
% One of the benefits of using this technology is having the capability of creating multi-hosts virtualized environments, which can be easily defined with the docker-compose files. This last function enables the use of dedicated backchannels for containers communication and the selection of the services needed to face outside networks. All these capabilities favor reproducibility, ease of deployment and seamless communication, which are key for a modular architecture like DISINFOX's. 

In particular, the Docker architecture defined for DISINFOX consists of (Figure \ref{fig:architecture}): 

\begin{figure}[h]
    \centering
    \includegraphics[width=1\linewidth]{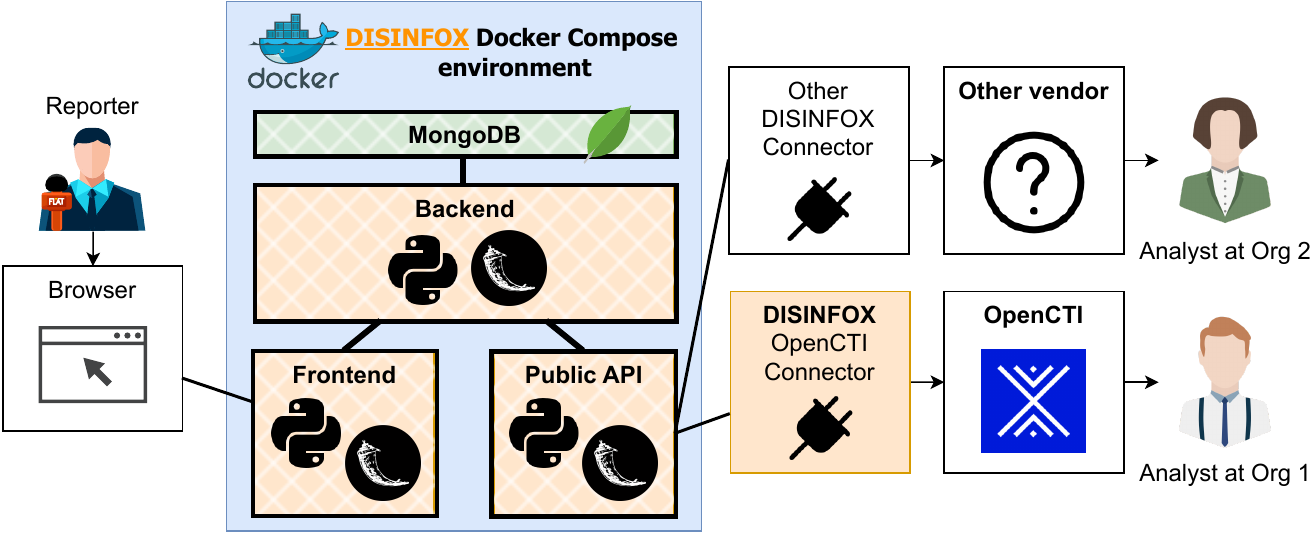}
    \caption{DISINFOX architecture}
    \label{fig:architecture}
\end{figure}

\subsubsection{Frontend}
A web-based interface designed for non-technical users enables them to share and view disinformation incidents easily. Built with Python 3 and Flask, it uses Jinja2 templates to render responsive and visually unified HTML pages using Bootstrap 5.3\footnote{\url{https://getbootstrap.com/docs/5.3/getting-started/introduction/}}. Also, Stixview\footnote{\url{https://github.com/traut/stixview}} was integrated to generate interactive STIX2 graphs, providing enhanced visualization of incidents. The frontend interacts with the backend to upload user-submitted incidents, display platform data, and manage user accounts.

\subsubsection{Backend REST API}
This component manages STIX2 objects and user data within the platform while interfacing securely with the data store.  
Developed with Python 3 and Flask, it provides a REST interface for handling STIX2 objects, enabling easy integration with future components and functionalities.  
Decoupling the backend from the frontend ensures the system remains agnostic to frontend technologies.  
The backend primarily sends STIX-formatted bundles to the frontend while ingesting and validating incidents submitted in the frontend.  
Using the STIX2 library, the backend transforms submitted data into well-formatted STIX2 objects and inserts them directly into the MongoDB collection.  
Additionally, this backend validates the public API requests and serves STIX2 objects to it for external CTI platforms.  

\subsubsection{Data Store}
A MongoDB database was selected for its native capability to store STIX2 objects.  
Various database types were evaluated, with SQL-based DBMSs discarded due to the extensive transformation required for STIX2 objects.  
Document-oriented DBMSs were preferred for their compatibility with JSON (the format used by STIX), offering flexibility and simplicity in handling the data.  
While graph databases could meet the requirements, their complexity and steep learning curve rendered them less suitable.  
Among document-oriented DBMSs, MongoDB was chosen for being open-source, providing robust Python library support, offering an official Docker image, and ranking as the most popular document database\footnote{\url{https://db-engines.com/en/ranking/document+store}}.

Although \DISINFOX is designed to function without preloaded data, allowing incidents to be added dynamically,  the open-source code provide a dataset of 118 incidents from a variety of sources. This dataset includes incidents from~\cite{fulde}, the DISARM repository\footnote{\url{https://github.com/DISARMFoundation/DISARMframeworks/blob/main/DISARM_MASTER_DATA/DISARM_DATA_MASTER.xlsx}}, and several new incidents introduced in this work.

\subsubsection{Public REST API}
This API, also built with Flask and Python 3 exposes endpoints for programmatic access to \DISINFOX’s incident repository managed by the backend, allowing CTI connectors and other software to retrieve data. Users must authenticate requests by including an API key, which is generated in the Profile section of the frontend interface.  

\subsubsection{DISINFOX OpenCTI Connector}

The publicly available\footnote{\url{https://github.com/CyberDataLab/opencti-connector-disinfox}} Python 3 connector for the OpenCTI platform serves as a proof of concept for demonstrating \DISINFOX’s interoperability. This connector retrieves new content from \DISINFOX and integrates it seamlessly into OpenCTI. Thanks to using STIX2 natively, no extra steps for the ingestion to OpenCTI are needed.

OpenCTI was chosen as the platform to build the connector and validate the interoperability of the platform due to several key factors. First, it is part of the technology stack for disinformation sharing agreed upon by the EU and the United States~\cite{ministerialttc}. Second, OpenCTI demonstrates a commitment to adapting its platform to better support disinformation management~\cite{filigranOpenCTIHelps}. Third, it is the most popular open-source platform capable of ingesting STIX2. Lastly, OpenCTI offers a comprehensive guide for building connectors and has strong Python library support through the \textit{ctipy} library.

\vspace{0.5cm}

While DISINFOX relies on all these modules for full functionality, only the frontend and the public REST API directly interact with external users,  
serving as the primary entry points to the platform.  

\subsection{Software functionalities} \label{sec:lifecycle} 

%\subsubsection{DISINFOX validation and incident lifecycle} \label{sec:lifecycle} 

\begin{figure}
    \centering
    \includegraphics[width=1\linewidth]{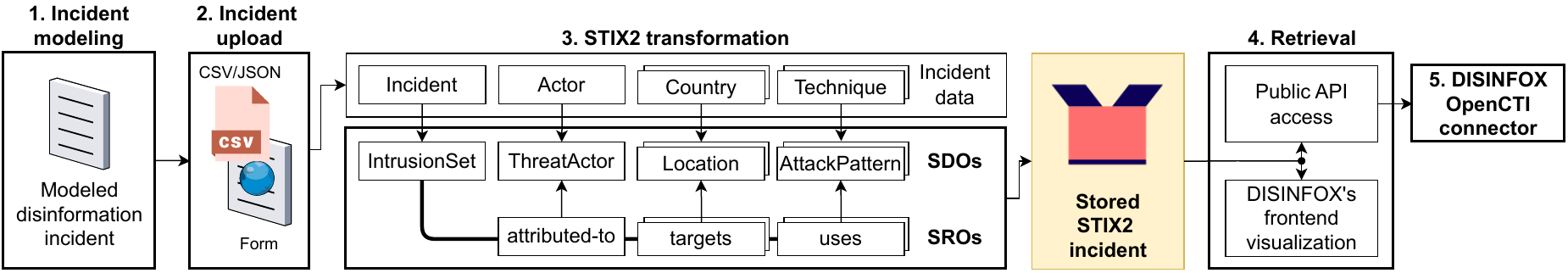}
    \caption{DISINFOX lifecycle}
    \label{fig:lifecycle}
\end{figure}

The following subsections detail how a disinformation incident is managed and shared within \DISINFOX. 
To illustrate the process, a use case related to the Ukraine war is referenced throughout. 
Figure \ref{fig:lifecycle} outlines the main steps in the lifecycle, from incident upload to ingestion by other CTI platforms. 
These steps were performed to generate 118 disinformation incidents from the ingestion of \DISINFOX's default dataset \footnote{\url{https://github.com/CyberDataLab/disinfox/blob/main/backend/data/merged_Foulde_DSRM_additions.csv}}.

\subsubsection{Incident modeling} \label{sec:modeling}

In this platform, the data model presented in our recent paper \cite{disinfoxcosec} is used to provide a simple, interoperable and structured way to categorize disinformation incidents thanks to DISARM TTPs.
Therefore, for a Reporter user to upload an incident to the platform, they must first identify a disinformation incident and recognize their TTPs.

In the following sections, we use the \textit{Bucha massacre} disinformation campaign \cite{buchadfrlab} as a real-world example to illustrate how incidents are managed within \DISINFOX. This incident, which took place in April 2022, contained at least 12 identified DISARM techniques employed by Russia against Ukraine.

\subsubsection{Incident upload}

\begin{figure}
    \centering
    \includegraphics[width=\textwidth]{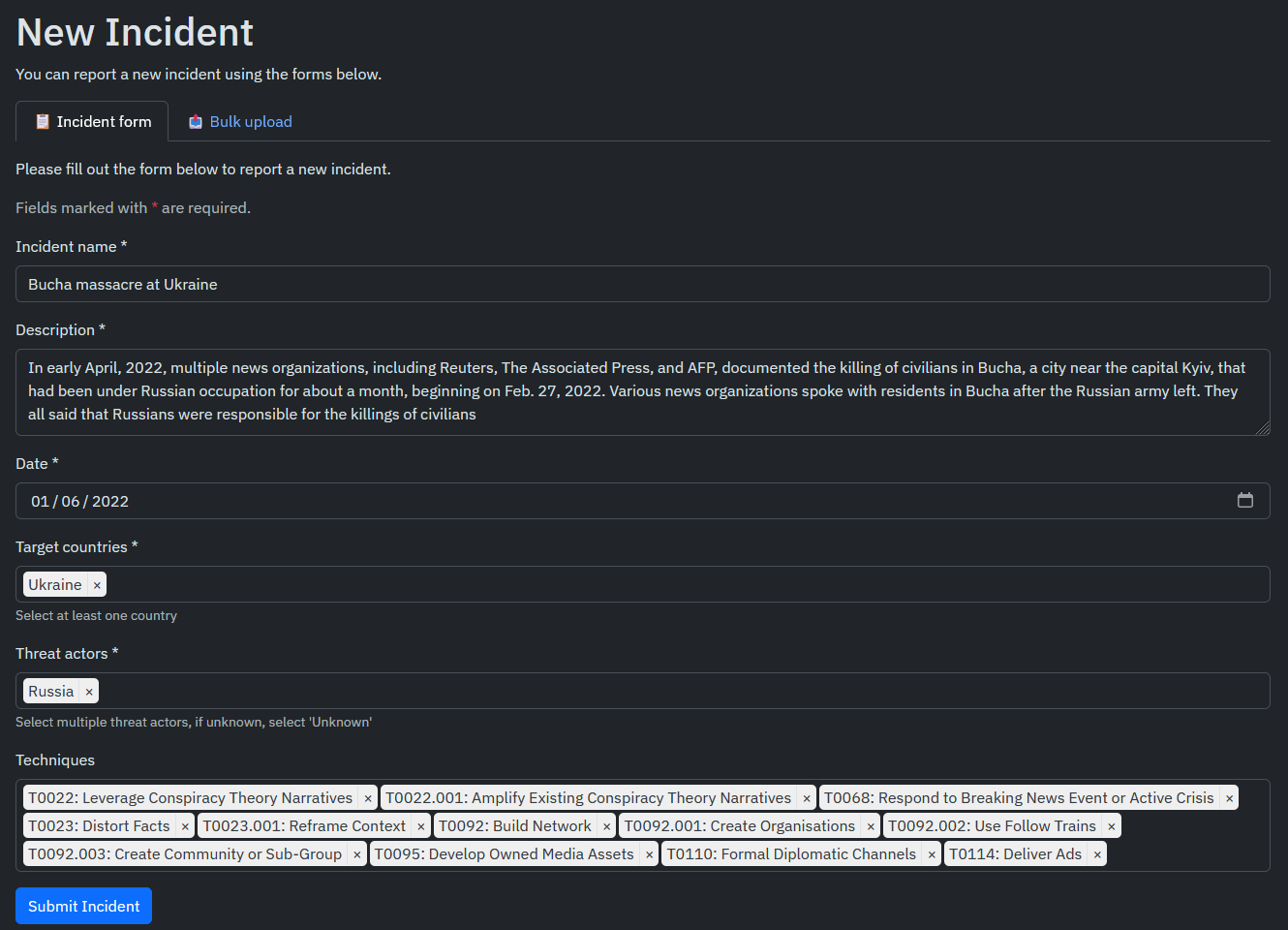}
    \caption{Manual individual upload form.}
    \label{fig:incidentupload}
\end{figure}

After a Reporter user identifies a disinformation incident and extracts the DISARM TTPs, it can upload it through \DISINFOX's frontend by using one of two methods: 

\begin{itemize}
    \item \textbf{Manual individual upload:} This is the simplest method for uploading a single identified incident. As illustrated in Figure \ref{fig:incidentupload}, the user needs to fill out a form with the following fields: incident name, description, date, target countries, threat actors, and identified DISARM techniques.
    \item \textbf{Bulk upload:} This method is ideal for importing a large set of disinformation incidents. The user can upload either a CSV file\footnote{The CSV file must follow a specific template based on the one used in this working paper~\cite{fulde}.} or a JSON file containing a STIX2 bundle with the incidents they wish to import. When using this method, the platform performs an intermediate transformation to format each individual incident, simplifying the creation of STIX2 objects.
\end{itemize}

The interactive form provides a user-friendly way for the Reporter to upload all the necessary information about a disinformation incident. The incident presented in Section \ref{sec:modeling} can be used as an example of how to fill out the form. 
Figure \ref{fig:incidentupload} illustrates the form fields filled with the required information for the incident. The title is entered as \textit{Bucha massacre at Ukraine}, while the description contains a summary of the source report. The date field is filled with \textit{April 1, 2022}, the date of the first evidence of disinformation. The target country is \textit{Ukraine}, as it was the target of the false claims. The threat actor is identified as \textit{Russia}, as noted in the source report. Finally, the DISARM techniques are listed according to those identified by the actions cited in the report.

Once incidents are uploaded using either method, the platform performs validation checks on the submitted data and transforms the incidents into individual STIX2 objects.

\subsubsection{Automated STIX2 transformation}

Once a disinformation incident is uploaded to \DISINFOX, the process of creating STIX2 objects from incident data is guided by the mapping established in \cite{disinfoxcosec}. The following steps indicate how STIX2 Domain Objects (SDOs) are built and then connected using STIX2 Relationship Objects (SROs) to represent a disinformation incident in compliance with the STIX2 standard in \DISINFOX:

\begin{enumerate}
    \item The uploaded disinformation incident data is used to generate individual SDOs, which are temporarily stored using Python's \textit{stix2} library. These include \texttt{IntrusionSet}, \texttt{ThreatActor}, and \texttt{Location} objects.

    % First, an \texttt{IntrusionSet} SDO is created using the title, description, and date provided in the form, which populates the \texttt{name}, \texttt{description}, and \texttt{first\_seen} properties of the object, respectively.  
    % Next, a \texttt{ThreatActor} SDO is generated using the threat actor names specified in the form.  
    % Finally, a \texttt{Location} SDO is created using the country names provided in the form.  
    
    % Using the incident example, the first three objects in Listing \ref{lst:stix2example} demonstrate how these properties are populated and aligned with the uploaded data.

    \item  All DISARM techniques are already represented as \texttt{AttackPattern} SDOs, pre-built and stored in the \texttt{DISARM.json} file\footnote{\url{https://github.com/DISARMFoundation/DISARMframeworks/blob/main/generated_files/DISARM_STIX/DISARM.json}} in STIX2 format.  
    The DISARM techniques selected in the form are iterated through and matched against their corresponding entries in the JSON file. For each matching technique, the JSON object is converted into a Python STIX2 object and temporarily stored.  
    
    % The fourth object in Listing \ref{lst:stix2example} \todo{aqui habia antes los listings de cada SDO/SRO, seguir?} illustrates how a DISARM technique identified in the incident is represented in STIX2 format. Note that the \texttt{created} date in the \texttt{AttackPattern} SDO reflects the last update of \texttt{DISARM.json}, not the upload date of the incident in \DISINFOX.
     
    \item SROs are generated to link the previously created SDOs, establishing relationships between the \texttt{IntrusionSet} and the \texttt{ThreatActor}, \texttt{Location}, and \texttt{AttackPattern} SDOs. These connections are represented using the \texttt{attributed-to}, \texttt{targets}, and \texttt{uses} relationship types, respectively.
    
    \item  All generated SDOs and SROs are inserted into the platform's database.
\end{enumerate}

The disinformation threat landscape is constructed from the STIX2 objects stored in the database, forming a structured and interoperable dataset for further analysis and sharing.

\subsubsection{Methods for incidents retrieval} \label{sec:retrieval}

The disinformation incidents stored in \DISINFOX can be queried in several ways, depending on the needs of the user:

\begin{figure}
    \centering
    \includegraphics[width=\linewidth]{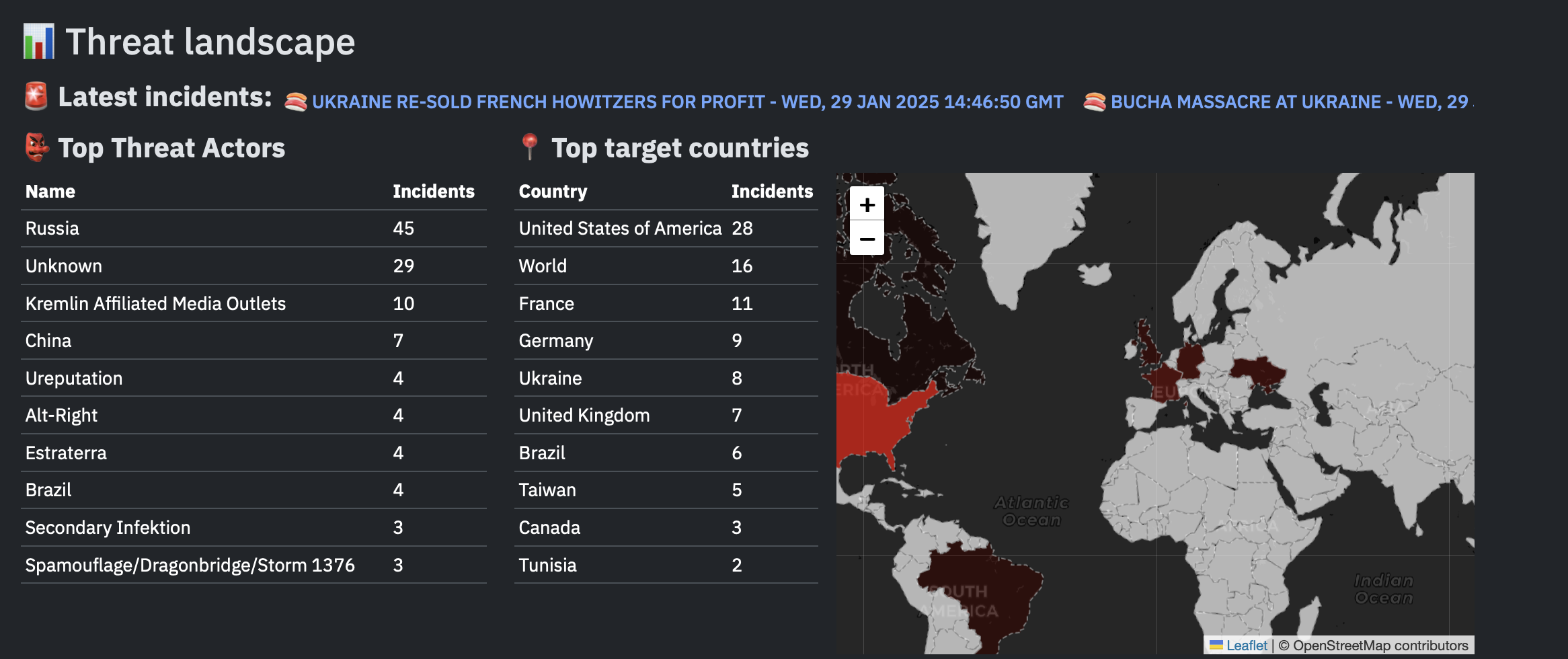}
    \caption{DISINFOX dashboard}
    \label{fig:dashboard}
\end{figure}

\begin{figure}
    \centering
    \includegraphics[width=1\linewidth]{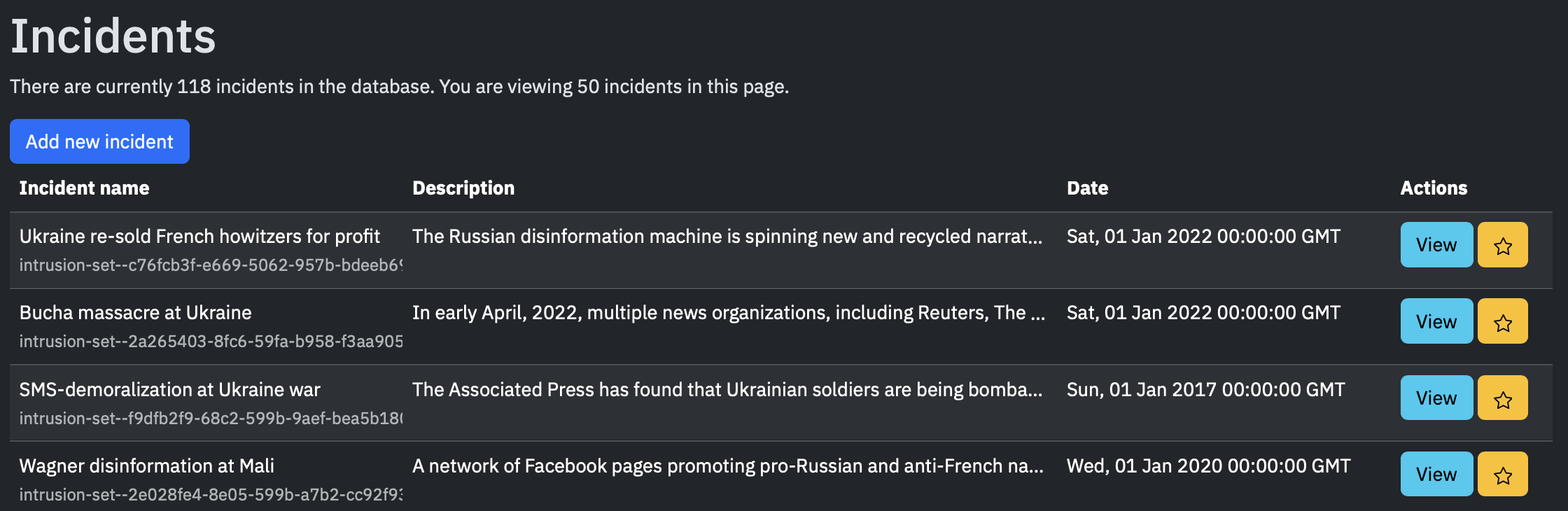}
    \caption{DISINFOX incident listing}
    \label{fig:listing}
\end{figure}

\begin{figure}
    \centering
    \includegraphics[width=\linewidth]{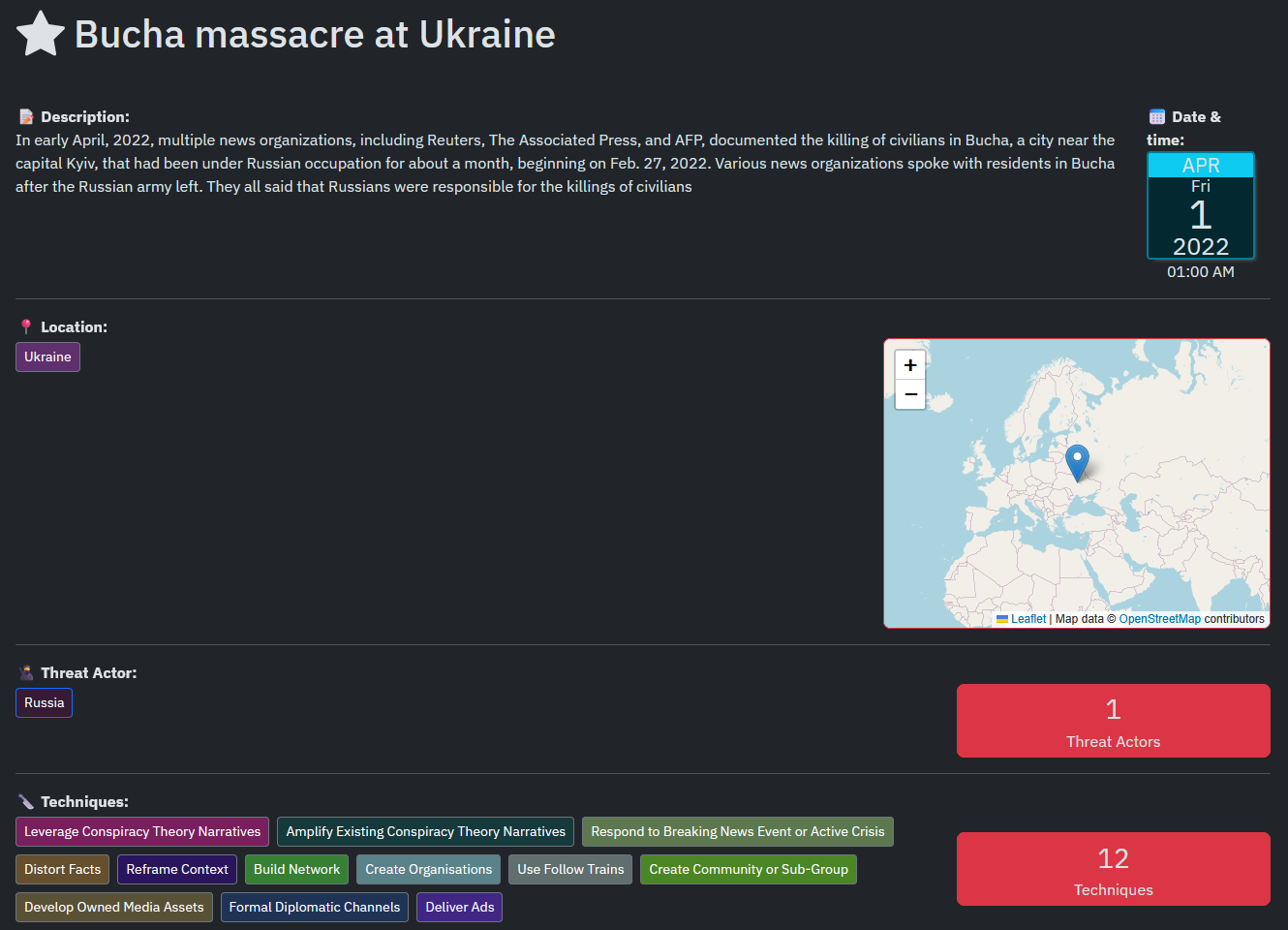}
    \includegraphics[width=\linewidth]{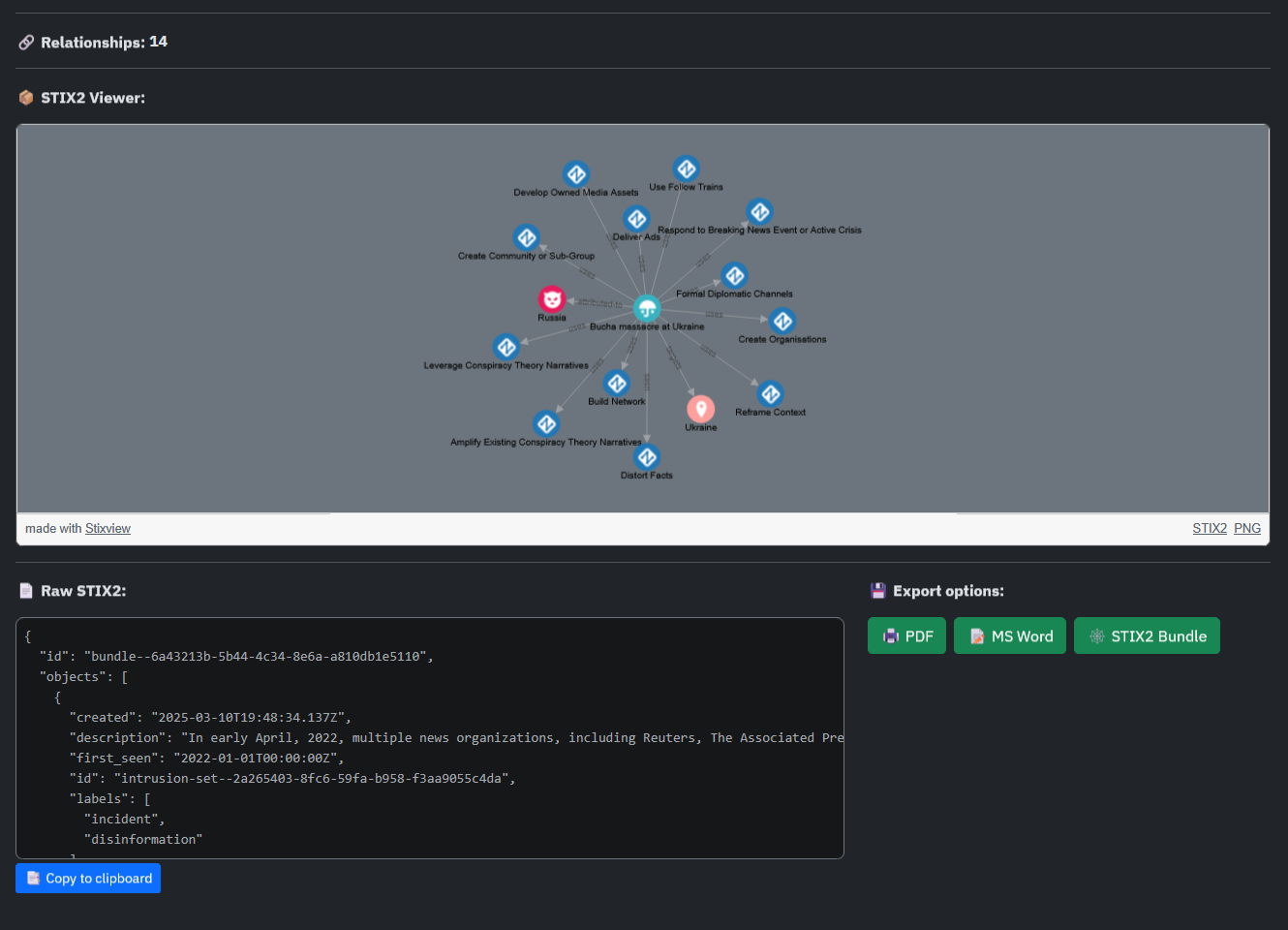}
    \caption{Visualization of a disinformation incident at the DISINFOX frontend web page}
    \label{fig:incident-detail}
\end{figure}

\begin{itemize}
    \item \textbf{For non-technical and casual users}, the most effective way of checking incidents is by looking at DISINFOX's frontend webpage.
    The first utility presented to the user is the dashboard (Figure \ref{fig:dashboard}), which offers a quick overview of the current disinformation landscape: the last disinformation incidents, the most active threat actors and the most attacked countries with an associated heatmap.
    In addition, the user can check the list (Figure \ref{fig:listing}), to visit any of the incidents stored on the platform. It shows the name, short description and date of the incidents. 
    Once the user has found an interesting incident, he can view its details to get more information from it (Figure \ref{fig:incident-detail}).
    All the information about the disinformation incident is shown graphically and intuitively: 
    name, full description, date, target countries with a map, actors, used techniques, a graph showing the STIX2 relationships of this incident
    and the raw STIX2 bundle that represents this incident.
    Additionally, users can generate a PDF or Word report with all the detailed information about the incident to export it to other media and 
    can select the incident as a favorite, so it can be easily found in its Profile.

    \item \textbf{Technical users and specialized CTI developers} have the option to use the Public API to query the platform. 
    Access to the API requires presenting an API key in the HTTP \texttt{Authorization} header, ensuring proper access monitoring and security. 
    To obtain an API key, developers must register on the platform and navigate to the API Key section in their Profile. 
    Once the API key is obtained, the Public API can be queried, as indicated in the messages between the connector and the Public API shown in Figure \ref{fig:connector-flow}. 
    
    The request to the \texttt{/incidents} endpoint should include the \texttt{newer\_than} parameter, 
    which takes an ISO 8601 datetime string with microsecond precision. 
    This parameter specifies the point in time from which the last edited STIX2 objects will be retrieved, 
    making it particularly useful for reducing traffic and retrieval times by fetching only new or updated information from the platform. If all the objects need to be retrieved, the epoch datetime can be used.
    
    This retrieval method allows developers to easily integrate incident data into their applications in a RESTful manner. Extending this functionality to support the ingestion of new incidents through the API is a goal for future development.
\end{itemize}

These two methods are essential to provide a useful way of retrieving incidents for two different use cases.

\subsubsection{Incident synchronization and visualization in OpenCTI}

As stated in Section \ref{sec:retrieval}, the Public API eases the work of incident retrieval for applications that want to use \DISINFOX's incidents, especially to connect it to other CTI solutions.

To demonstrate this, the proof-of-concept \DISINFOX connector for OpenCTI 6.4.2\footnote{\url{https://github.com/CyberDataLab/opencti-connector-disinfox}} was developed. Although the \DISINFOX connector can be used standalone with an OpenCTI installation, it is recommended to first install the DISARM connector\footnote{\url{https://github.com/OpenCTI-Platform/connectors/tree/master/external-import/disarm-framework}}. The DISARM connector not only inserts all \texttt{AttackPattern} SDOs from DISARM into OpenCTI but also provides the DISARM matrix and other additional objects that enhance the utility of the \DISINFOX connector. This allows the \DISINFOX incidents shared with OpenCTI to be analyzed using the matrix, complementing all the other visualization options available in OpenCTI.

\begin{figure}
    \centering
    \includegraphics[width=\linewidth]{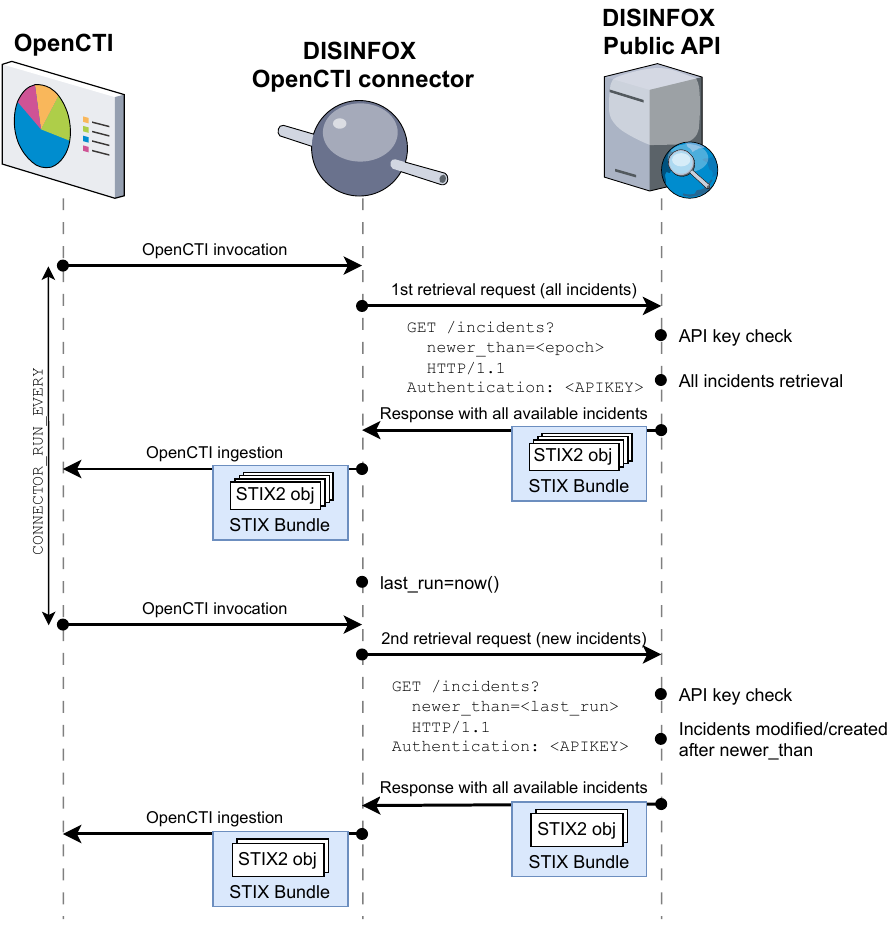}
    \caption{DISINFOX's proof-of-concept OpenCTI connector messages}
    \label{fig:connector-flow}
\end{figure}

As Figure \ref{fig:connector-flow} shows, this connector works in a very simple way thanks to using STIX2 natively:
\begin{enumerate}
    \item The OpenCTI platform registers the connector and performs the first run of \DISINFOX's connector.
    \item \DISINFOX connector sends a request to \DISINFOX Public API with the \texttt{newer\_than} 
    parameter set with the epoch timestamp, as this is the first run, and all the incidents need to be retrieved.
    It also includes an \texttt{Authorization} header with the API key that the user have included in the \texttt{.env} file,
    previously obtained through its \DISINFOX's profile.
    \item \DISINFOX Public API checks the API key in the request headers. If it is valid, it starts 
    retrieving all the incidents from the backend and sends them back to the \DISINFOX connector as a response.
    The body of this response will contain all the STIX2 objects representing all the incidents uploaded
    to the platform.
    \item \DISINFOX connector inserts the STIX2 objects from the API response to OpenCTI without
    any extra transformation.
    \item The last operations are repeated just by changing the \texttt{newer\_than value}, which now will be set to the last time that the connector was set.
    The next call to the connector will be made depending on the time set in the \texttt{CONNECTOR\_RUN\_EVERY} parameter set in the installation of the connector to OpenCTI.
\end{enumerate}

Now, all SDOs and SROs are stored in OpenCTI.
A listing of all the ingested disinformation incidents can be easily seen in the \textit{Threats \textgreater\ Intrusion Set} section.

The presented use case can be used as an example to see the analysis that can be done in the OpenCTI platform.
Apart from the \textit{Overview} section that shows a summary of the properties (name, description, first seen date, etc.),
the \textit{Knowledge} tab of the Ukrainian incident offers much more interesting data.

\begin{figure}
    \centering
    \includegraphics[width=\linewidth]{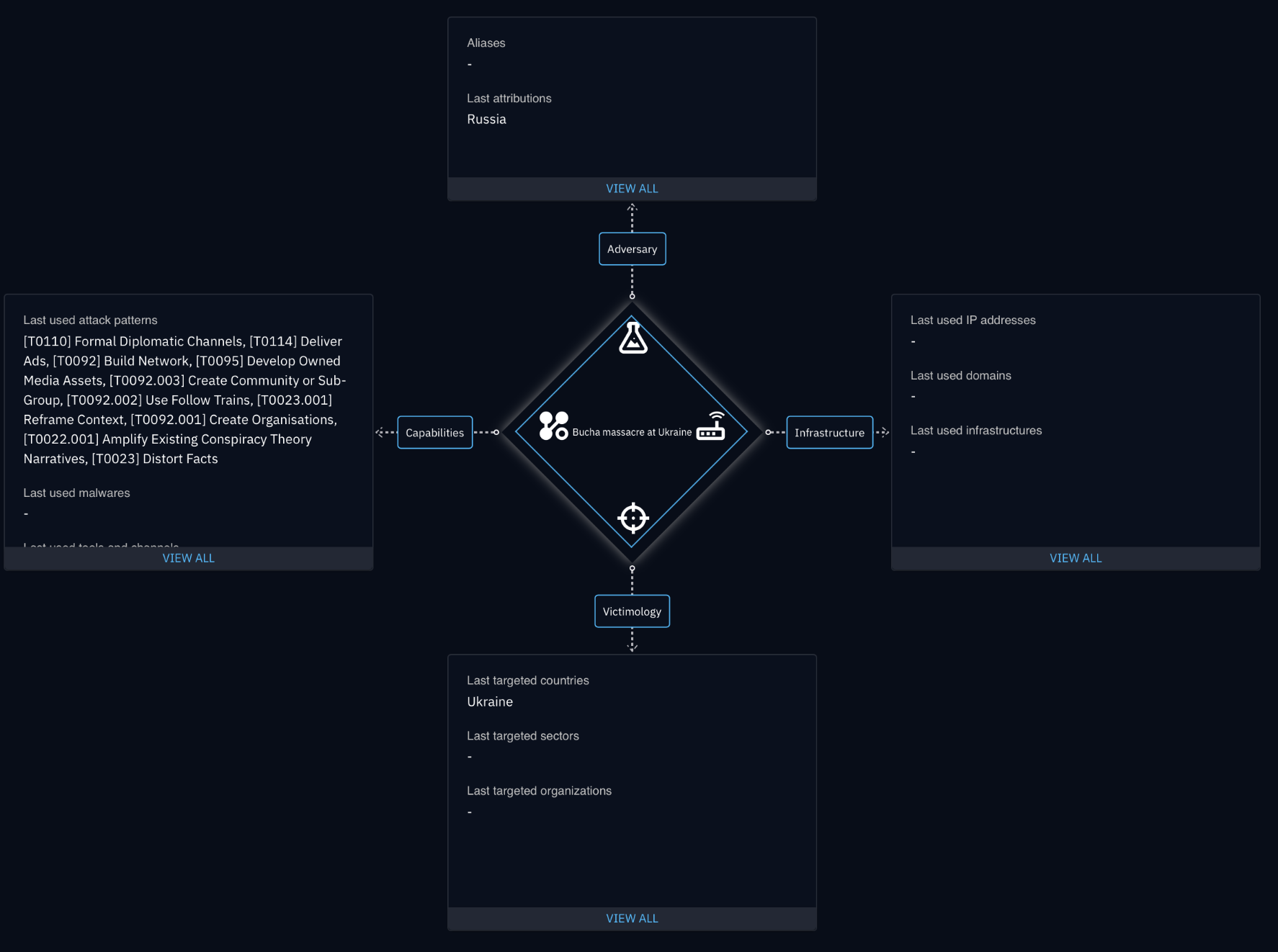}
    \includegraphics[width=\linewidth]{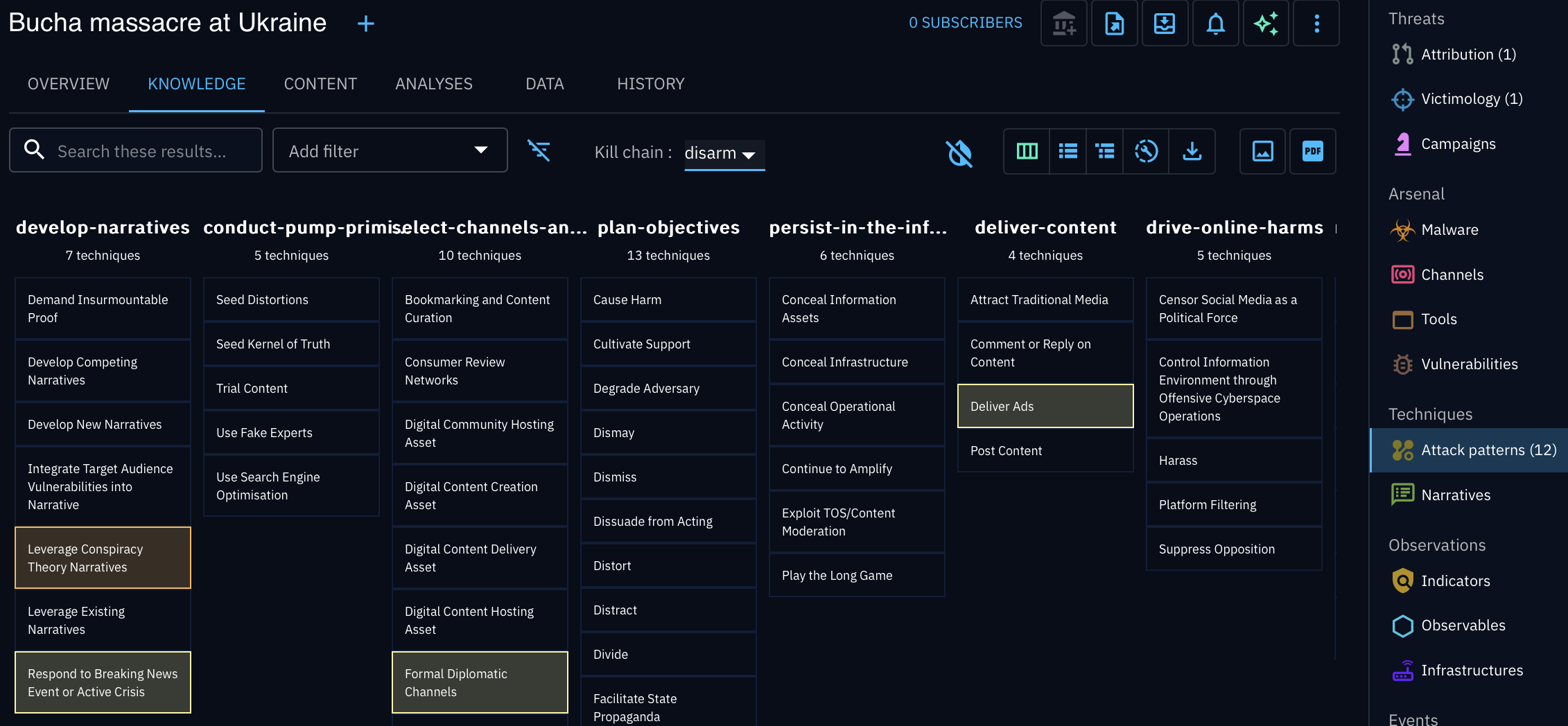}
    \caption{OpenCTI \textit{Knowledge} tab in the page of the modeled intrusion set}
    \label{fig:octi-knowledge}
\end{figure}

The first picture of Figure \ref{fig:octi-knowledge} shows the \textit{Diamond} graph that summarizes the relationships of the intrusion set in 4 dimensions:
\textit{Adversary}, where we find Russia as the threat actor; \textit{Capabilities}, where attack patterns (DISARM techniques) such as
 \textit{Formal Diplomatic Channels} or \textit{Deliver Ads} can be directly found; \textit{Victimology}, where Ukraine is set as the 
target of this intrusion set; and \textit{Infraestructure}, which is unused.

If the \textit{VIEW ALL} button in the \textit{Capabilities} frame or the \textit{Attack patterns} button in the right bar is selected, 
OpenCTI displays the view in the second picture of Figure \ref{fig:octi-knowledge}.
This is the matrix view, which shows the used attack patterns in the matrix model that is selected, in this case, the DISARM matrix,
which has been installed thanks to the DISARM connector.
Notice how all the attack patterns used in the Ukrainian incident are highlighted under their corresponding tactic in the DISARM matrix.

These are just examples of the possibilities of using OpenCTI to manage disinformation incidents, but other
actions such as Cyber Kill Chain analysis or correlation with other incidents by taking into account its common DISARM techniques or target locations can be achieved. Overall, disinformation analysts can embed this connector into their workflow to monitor, correlate and asses disinformation incidents with a potentially shared view with other cybersecurity incidents, providing a rich picture of the current picture of the threat landscape.

\section{Conclusions}

This work introduces \DISINFOX, an open-source threat intelligence exchange platform for managing and distributing disinformation incident data. \DISINFOX facilitates the structured and interoperable reporting, visualization, and analysis of disinformation incidents by leveraging DISARM TTPs and a custom STIX2-based data model. A use case illustrates how these functionalities operate in practice.  

Disinformation incidents can be easily uploaded through the web-based frontend, which supports automated detection of TTPs in complementary PDF reports. The platform provides an interactive listing of incidents, allowing users to explore detailed information, including descriptions, affected countries visualized on a map, associated DISARM TTPs, and a STIX2 graph representation of the incident. Additionally, incidents can be exported in more readable formats such as PDF or Word, alongside the original STIX2 Bundle.  

The interoperability of \DISINFOX ensures seamless integration with other CTI solutions through its Public API, enabling the ingestion of structured disinformation incidents into mature cybersecurity platforms. This interoperability enhances correlation and investigation capabilities by allowing analysts to link disinformation campaigns with traditional cybersecurity threats, reflecting real-world attack scenarios.  

To validate this approach, a proof-of-concept \DISINFOX connector for OpenCTI was developed, successfully ingesting over 100 modeled disinformation incidents from various sources. The technology stack adopted (\textit{DISARM + STIX2.1 + OpenCTI}) aligns with the strategy jointly agreed upon by the EU and the US for addressing Foreign Information Manipulation and Interference (FIMI), as outlined in the \textit{EU-US Trade and Technology Council’s} fourth ministerial meeting~\cite{ministerialttc}.

\section*{Acknowledgements}

This study was partially funded by (a) the strategic project ``Development of Professionals and Researchers in Cybersecurity, Cyberdefense and Data Science (CDL-TALENTUM)" from i) the Spanish National Institute of Cybersecurity (INCIBE) and ii) by the Recovery, Transformation and Resilience Plan, Next Generation EU, and (b) by a ``Juan de la Cierva'' Postdoctoral Fellowship (JDC2023-051658-I) funded by the i) Spanish Ministry of Science, Innovation and Universities (MCIU), ii) by the Spanish State Research Agency (AEI/10.13039/501100011033) and iii) by the European Social Fund Plus (FSE+).

%% The Appendices part is started with the command \appendix;
%% appendix sections are then done as normal sections
%% \appendix

%% \section{}
%% \label{}

%% References:
%% If you have bibdatabase file and want bibtex to generate the
%% bibitems, please use
%%
%%  \bibliographystyle{elsarticle-num} 
%%  \bibliography{<your bibdatabase>}

%% else use the following coding to input the bibitems directly in the
%% TeX file.

\bibliographystyle{elsarticle-num} 
\bibliography{mybib}

\end{document}